\documentclass[twocolumn,pre]{revtex4}
\usepackage{graphics,graphicx,epsfig}
\usepackage{epsf,epstopdf,wrapfig}
\usepackage{amssymb,amsfonts,amsmath}
\usepackage{graphicx,pstricks,epsfig}
\include{epsf}

\newcommand{\<}{\langle}
\renewcommand{\>}{\rangle}
\newcommand{\degree}{{}^{\rm o}}

\begin{document}
\title{Steps in the bacterial flagellar motor}

\author{Thierry Mora}

\address{Joseph Henry Laboratories of Physics and Lewis-Sigler Institute for Integrative Genomics,
Princeton University, Princeton, New Jersey, USA}

\author{Howard Yu}

\address{Joseph Henry Laboratories of Physics,
Princeton University, Princeton, New Jersey, USA}

\author{Yoshiyuki Sowa}

\address{Clarendon Laboratory, Department of Physics, University of Oxford, Oxford, UK} 

\author{Ned S. Wingreen}

\address{Department of Molecular Biology,
Princeton University, Princeton, New Jersey, USA}


\begin{abstract}
The bacterial flagellar motor is a highly efficient rotary machine used by many bacteria to propel themselves. It has recently been shown that at low speeds its rotation proceeds in steps [Sowa {\em et al.} (2005) {\em Nature} 437, 916--919]. Here we propose a simple physical model that accounts for this stepping behavior as a random walk in a tilted corrugated potential that combines torque and contact forces.
We argue that the absolute angular position of the rotor is crucial for understanding step properties, and show this hypothesis to be consistent with the available data, in particular the observation that backward steps are smaller on average than forward steps. Our model also predicts a sublinear torque-speed relationship at low torque, and a peak in rotor diffusion as a function of torque.

\end{abstract}

\maketitle

%
%


Bacteria swim by virtue of tiny rotary motors that drive rotation of helical flagella. These motors are powered by a transmembrane proton (or Na${}^{+}$) flux which is converted into torque.
However, little is known about the detailed mechanisms of energy conversion, or torque generation. Recently, a new result has provided direct insight into motor operation \cite{Sowa:2005p8}: at low speeds, the bacterial flagellar motor proceeds by steps.
This stepping is stochastic in nature, as manifested by the occurrence of occasional backward steps even for motors locked in one rotation direction.
What is the origin of motor steps and how can these steps be reconciled with the near perfect efficiency of the motor observed at low speeds \cite{MEISTER:1987p19}? We argue that steps, including backward steps, are an inevitable consequence of the physical structure of the motor---a stator driving a ``bumpy'' rotor through a viscous medium.

\begin{figure*}
\includegraphics[width=\linewidth]{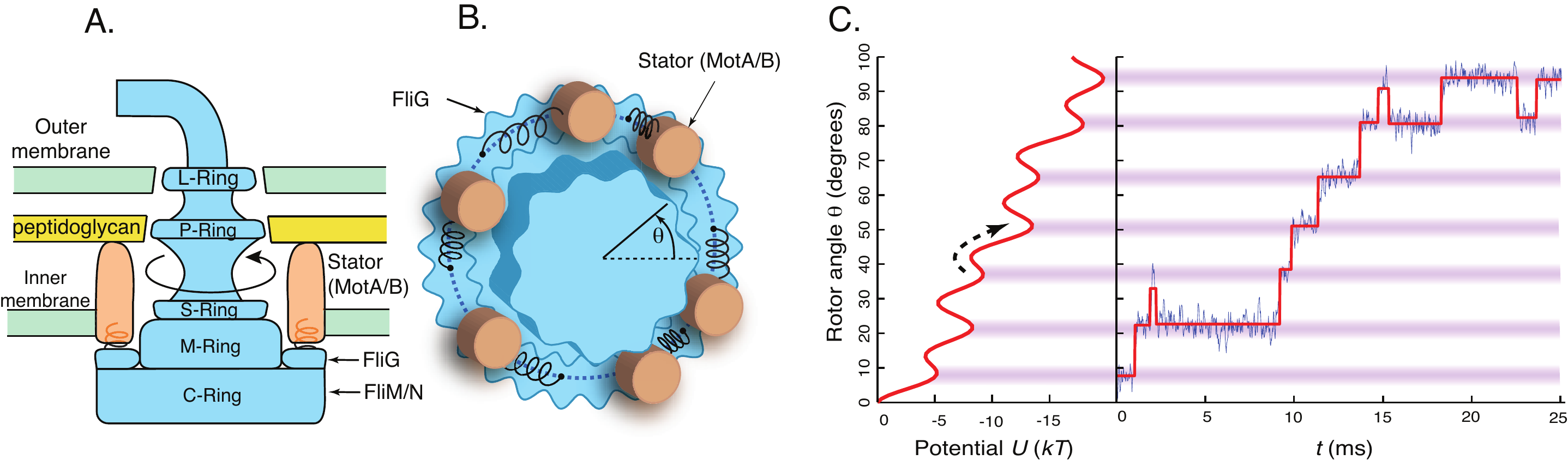}
\caption{{\bf Model for stepping of the flagellar motor. A. }Side view of the flagellar motor. {\bf B.} Top view of the motor highlighting the model's essential ingredients. The passage of H${}^{+}$ across the inner membrane causes the stretching of protein ``springs'' which link the fixed stator complexes (MotA/B) to the rotor (FliG, etc.). The stretched springs apply a torque to the rotor. Contact forces between the stators and the rotor also produce a potential of interaction, which is approximately 26-fold periodic due to the 26 FliG subunits. {\bf C.} Left: Rotation of the rotor as a whole corresponds to a viscously damped random walk in a tilted corrugated potential $U(\theta)$ arising from the combined torque and contact potential. Right: Example of a trace generated by the model (blue) and the inferred steps (red) between local potential wells (shown with purple shading).
}
\label{fig:modelmakessteps}
\end{figure*}

In response to chemotactic signals, flagellar motors switch from counterclockwise to clockwise rotation causing cells to tumble or change directions. In {\em E. coli}, the basic mechanism of torque generation appears to be the same for both directions of motor rotation \cite{BlairBerg88}.
Torque is generated by the passage of H${}^{+}$ ions (or in some organisms Na${}^{+}$ ions) through the cytoplasmic membrane.
As shown schematically in Fig.~\ref{fig:modelmakessteps}A, torque is applied to the rotor, including the flagellum, by the stator, which is comprised of independent torque-generating units (MotA/B complexes) anchored to the peptidoglycan cell wall. 
The exact number of torque-generating units can vary from motor to motor, with the maximum estimated to be at least 11 \cite{Reid:2006p17}.
The rotor includes 26 circularly arrayed FliG proteins that contact the MotA/B complexes. The torque-speed relation of the motor has been measured under a range of conditions \cite{Berg:1995p18,Chen:2000p5,Ryu:2000p12,Inoue:2008p11}. The maximum torque in the high load, low speed regime tracks the electrochemical potential difference or proton motive force (PMF) across the membrane, and the motor operates with nearly perfect efficiency \cite{MEISTER:1987p19}. Whereas torque and efficiency fall off at high speeds, proton flux and motor rotation are always strongly coupled with $\approx\ 120$ protons passing through the membrane per MotA/B unit per rotation \cite{MEISTER:1987p19}.

Recent experiments, where rotation was measured by attaching a polystyrene bead to a flagellar stump driven by a counterclockwise-locked Na${}^{+}$-powered chimaeric motor at low speeds (low Na${}^{+}$-electrochemical-potential difference and low stator number), revealed that the motor proceeds by steps \cite{Sowa:2005p8}.
The steps have average size $\approx 13.8\degree$, which corresponds to 26 steps per rotation, exactly the number of copies of FliG around the rotor.
Occasional backward steps are observed and, interestingly, these are smaller on average than forward steps ($10.9\degree$ versus $13.8\degree$).
These observations, as well as the stepping mechanism itself, have so far remained unexplained. 
It has been suggested that stepping is caused by the stochastic passage of ions.
However, as pointed out in \cite{Sowa:2005p8}, the energy provided by passage of a single ion can only move the rotor attached to a 1 $\mu$m polystyrene bead by $5\degree$, much less than the typical observed step size.

Here we propose a simple physical model to explain stepping:
the stator applies nearly constant torque to the rotor, but, at the same time, contact forces on the rotor produce a potential and therefore an additional torque with approximately the 26-fold periodicity of FliG.
Our model naturally accounts for the existence of backward steps, as well as the discrepancy between forward and backward step sizes, and also predicts that step statistics depend on the absolute position of the rotor around the circle.
Our predictions are found to be consistent with the available data, including angular diffusion of the motor \cite{Samuel:1996p40}, and suggest how steps could be used to study the physical structure of the motor.
A novel testable prediction is that the torque-speed relation will become sublinear at very low torques.

\section*{Model}
Our model for stepping relies on two main assumptions: constant or nearly constant torque between stator and rotor and an approximately 26-fold periodic contact potential.

\noindent (1)
All torque-generating units apply torque simultaneously and additively. Following the model of Meister {\em et al.} \cite{MEISTER:1989p20}, we assume that each MotA/B complex acts as a set of protein springs that reversibly store the energy available from H${}^{+}$ (or Na${}^{+}$) translocations (see Fig.~\ref{fig:modelmakessteps}B). 
The protein springs are attached to fixed sites of the rotor circumference. When an H${}^{+}$ passes through the membrane, it causes a spring to detach from its attachment site, stretch, and reattach to the next site.
At stall, all springs are maximally stretched, such that the PMF matches the energy necessary to stretch a spring to its next site. At low speeds, the rotor moves and springs relax, but these are quickly restretched by H${}^{+}$ passage, so that the system remains in quasi-equilibrium with the torque set by the PMF.
Spring stretching may vary slightly among units, but since there are several motor units, the instantaneous torque self-averages and is nearly constant in time. Under this scenario, steps cannot be explained at the level of a single pair of MotA/B and FliG subunits, but must arise at the global level of the rotor-stator interaction.

\noindent (2)
There are contact forces between the stator and the rotor. These forces may be caused by contact between the MotA/B stator units and FliG proteins, but also possibly by contact with FliF (M-S ring), FlgG, FlgH or FlgI proteins (distal rod, L, and P rings, respectively) each of which forms a circle of 26 copies. There may be other periodicities to the contact forces as well, arising from the filament and the hook, which are $11$-fold periodic, from FlgK and FlgL (hook-filament junction, 11 copies each), FlgB, FlgC and FlgF (proximal rod between L and P rings, 6 copies each) and FliE (Rod-MS-ring junction, 9 copies).
We assume that a 26-fold periodicity is dominant, in agreement with experimental observations.
We therefore collect all contact forces in a potential $V(\theta)$ which we suppose to be approximately 26-fold periodic (Figure \ref{fig:modelmakessteps}C).

Since the motor operates at the molecular scale, its rotation is intrinsically stochastic as it is subject to random thermal fluctuations. Under the combined influence of the applied torque, the contact potential, and thermal fluctuations the rotor performs a continuous random walk in a tilted, approximately periodic potential, which we model by the following Langevin equation:
\begin{equation}\label{eq:langevin}
\frac{d \theta}{d t}=-\frac{1}{\nu}\frac{\partial U}{\partial{\theta}}+\xi(t),
\end{equation}
where $\nu$ is the drag coefficient, $\tau$ the total torque exerted by the stator, and where the potential $U(\theta)$ includes both the torque and the contact potential:
\begin{equation}\label{eq:langevin2}
\begin{array}{rccc}
U(\theta)=&V(\theta)&-&\tau\theta.\\
&\textrm{(contact potential)}&~&\textrm{(torque)}
~
\end{array}
\end{equation}
The term $\xi(t)$ represents Gaussian white noise and accounts for thermal fluctuations: $\<\xi(t)\xi(t')\>=2D\delta(t-t')$, where $D$ is the rotor diffusion coefficient, related to the temperature and the drag coefficient via Einstein's relation: $D=kT/\nu$.
In experiments, a load (usually a polystyrene bead) is attached to a flagellar stump and this load is largely responsible for the drag. For simplicity, we assume that linkage between motor and load is instantaneous, as the relaxation 
is rapid compared to the typical stepping time (see Discussion).

\begin{figure}
\includegraphics[width=\linewidth]{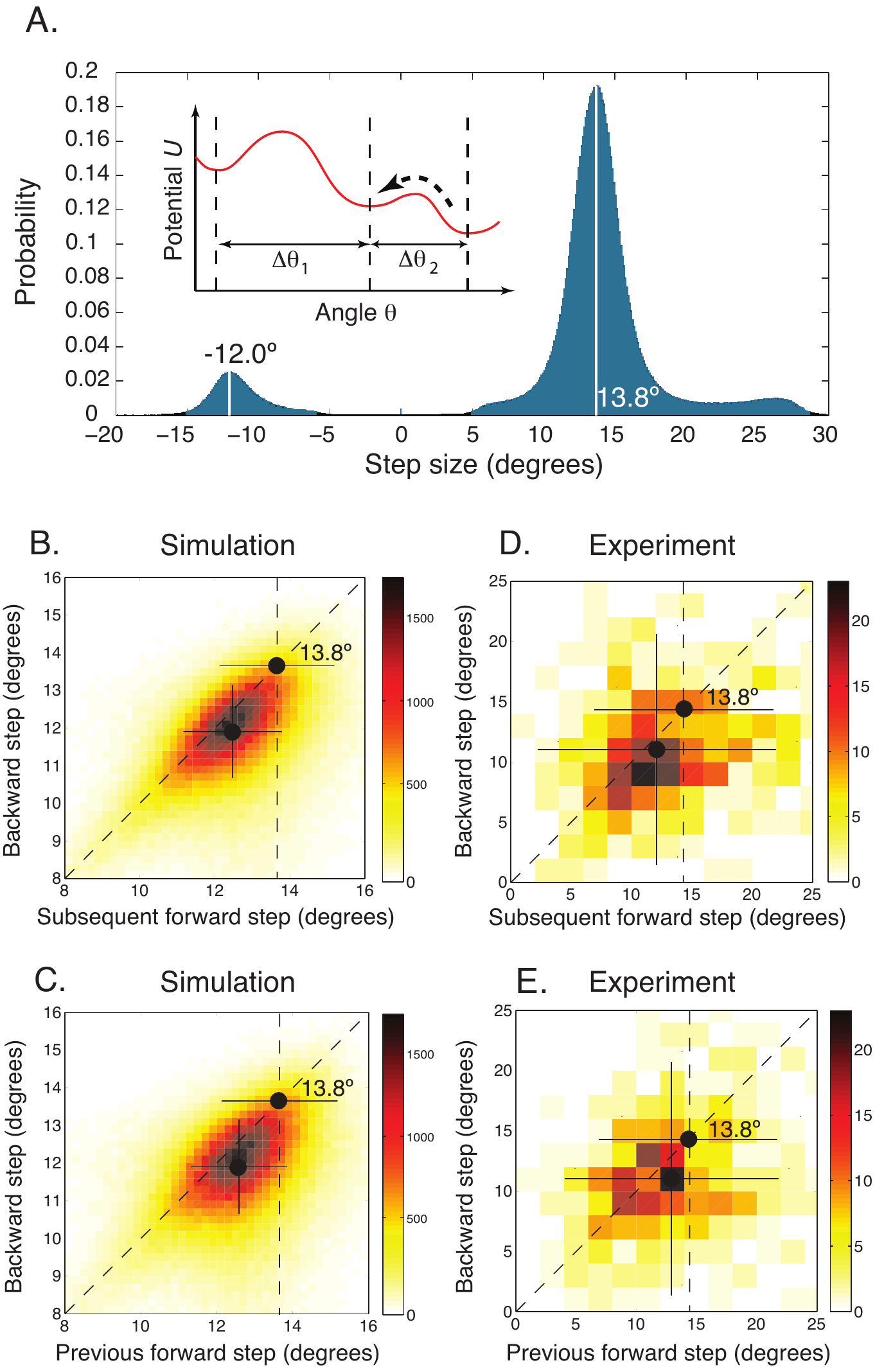}
\caption{{\bf Backward steps are smaller than forward steps. A.} Probability distribution of step sizes using a approximately 26-fold periodic potential (see main text), showing that backward steps are on average smaller than forward steps, in agreement with experiment. Inset: Backward steps rely on low barriers, which occur preferentially where angular steps sizes ({\em e.g.} $\Delta\theta_2$) are small. {\bf B-E.} Forward steps immediately following or preceding backward steps are found to be smaller on average ($12.0^{\rm o}$ in the experiment, $12.4^{\rm o}$ in the model) than the mean of all forward steps ($13.8^{\rm o}$ in model and experiment, black dots). Crosses denote mean and standard deviation of backward and subsequent or previous forward steps, while black dots and horizontal lines give the mean (13.8$\degree$) and standard deviations of all forward steps. Note change of scale between simulation and experiment. The experimental data are from \cite{Sowa:2005p8}.
}
\label{fig:backstepsaresmaller}
\end{figure}

\section*{Results}

\medskip\noindent\textbf{Steps are barrier-crossing events.} 
Numerical simulation of the model (Eq.~\ref{eq:langevin}) shows that rotation proceeds in steps (Fig.~\ref{fig:modelmakessteps}C). These steps correspond to jumps between adjacent wells of the tilted potential. Jumps/steps are possible thanks to thermal fluctuations which drive the system out of energy minima; without these fluctuations, the system would remain stuck in one well forever. Steps therefore correspond to crossings of the energy barriers separating wells. According to the Arrhenius law, the average time to cross a barrier increases exponentially with the barrier height. Because of the tilt induced by the torque, steps in the forward direction correspond to lower energy barriers than steps in the backward direction ({\em cf.} Fig.~\ref{fig:modelmakessteps}C). Forward steps are therefore more likely to occur than backward steps, so that on average the motor moves forward. 

\medskip\noindent\textbf{Backward steps are smaller than forward steps.}
To further investigate stepping in our model, we
wrote a step detector algorithm similar to that described in \cite{Sowa:2005p8}, and applied it to
a simulation of $\approx 2\cdot 10^5$ rotations (see Methods).
Fig.~\ref{fig:backstepsaresmaller}A shows the histogram of step sizes.
As in the experiment, we find that backward steps are smaller on average than forward steps: the mean forward step is $\approx 13.8\degree$ ($\approx 360\degree / 26$), against $12.0\degree$ for backward steps. (The precise values vary with the particular choice of potential and torque, but the mean step size is always larger for forward than backward steps.)

Recognizing steps as barrier crossing events allows us to readily explain the difference between average forward and backward steps sizes using a simple intuitive argument, as illustrated in the inset to Fig.~\ref{fig:backstepsaresmaller}A.
Backward steps occur infrequently because the energy barriers for these steps are higher than for forward steps. By contrast, all forward steps must occur as the motor moves forward, regardless of the heights of energy barriers. In addition, as shown in the inset to Fig.~\ref{fig:backstepsaresmaller}A, barrier heights and step sizes tend to be positively correlated. Roughly speaking, higher barriers extend over longer ranges. 
(We confirmed the generality of this correlation by simulating motion for many approximately 26-fold periodic potentials $V(\theta)$, with the results shown in Fig.~S1.)
Therefore, backward steps occur mostly over the lower barriers, and lower barriers correspond to smaller step sizes. This implies that backward steps are on average smaller than forward steps. Note that this argument relies on the fact that the barriers are not all identical.

To test the scenario proposed above, we asked whether the size of forward steps immediately preceding or following backward steps differ from the average of $13.8\degree$. According to our picture, the barrier crossed by these forward steps should be the same as the one crossed by the backward step immediately preceding or following, implying a small barrier and therefore a small forward step size.
Fig.~\ref{fig:backstepsaresmaller}B-E shows that indeed forward steps preceding of following a backward step are smaller on average, $12.0\degree$ in the experiment, $12.4\degree$ in our simulation, than the mean forward step of $13.8\degree$. (Note that even for forward steps over the same barrier, backward steps are still slightly smaller on average in both experiment and simulation. This suggests that the step detection algorithm has a small systematic bias---see supporting information (SI) Text.)

\begin{figure}
\includegraphics[width=\linewidth]{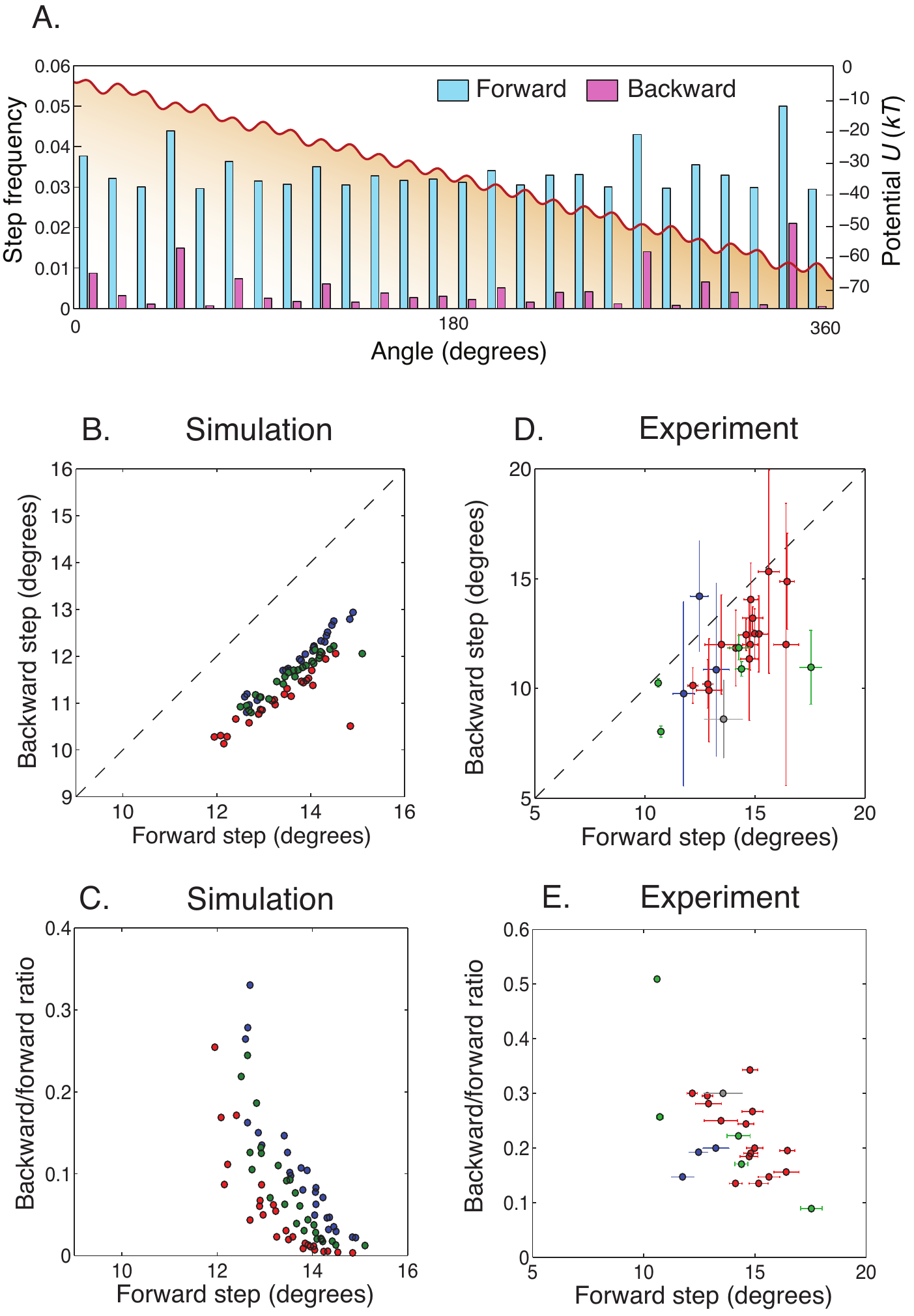}
\caption{{\bf Absolute position of the rotor matters. A.} A typical approximately 26-fold periodic tilted potential $U$ (red line) used in our simulations. For each of the 26 barriers in $U$, we show the frequencies of forward and backward steps across that barrier obtained analytically from first-passage theory (see Methods). Backward steps occur much more frequently at low barriers.
{\bf B.} Average backward and forward step sizes for each of the 26 barriers around the circle. Each color corresponds to a simulation with a different choice of the potential $U$.
{\bf C.} The ratio of backward over forward step counts for a given barrier decreases with the average forward step size (colors as in B). {\bf D and E.} Same as B and C, but with experimental data \cite{Sowa:2005p8}. Each color corresponds to a different cell. For each cell and each position around the circle, we show a data point only if there were at least 10 backward steps.
}
\label{fig:positionmatters}
\end{figure}

\medskip\noindent\textbf{Importance of absolute position of the rotor.}
The difference between forward and backward step sizes relies on the 26 barriers around the circle not all being identical. These heterogeneities may exist because the potential $V(\theta)$ is only approximately $26$-fold periodic and contains other periodicities as well arising from the filament, hook, or other parts of the rotor.  In any event, an essential prediction of our model is that step sizes and backward step frequencies will depend strongly on absolute position ({\em i.e.} modulo $360\degree$), reflecting the fixed contact potential $V(\theta)$.
We now examine how step frequencies and sizes depend on the properties of particular barriers, specified by the position of the rotor around the circle, and how this can tell us something about the detailed nature of contact forces.

According to our model, backward steps should be much more likely to occur at low barriers. In contrast, where forward steps occur should be much less sensitive to barrier heights. 
This follows simply because for each complete rotation the number of forward steps over any barrier is one plus the number of backward steps over that same barrier. Therefore, as long as backward steps are rare, the average number of forward steps over each barrier will be close to one per rotation and therefore the frequencies of forward steps will be similar for all barriers.
This is illustrated in Fig.~\ref{fig:positionmatters}A, which presents average step frequencies for each of the 26 barriers of a particular potential (chosen to be the same as in Fig.~\ref{fig:backstepsaresmaller}), as calculated from first-passage theory (see Methods).
As expected, there is considerable variation among barriers in backward-step frequencies, but much less variation in forward-step frequencies.

To relate average step sizes to absolute position, we examined the sizes of backward and forward steps for each of the $26$ barriers. To properly assign steps to barriers,
we sorted steps into $26$ equal bins according to the angular position of the rotor when the step occurred, and calculated the average backward and forward step sizes in each bin. We applied this procedure to simulations of $\sim 10^5$ rotations generated with three different potentials $U(\theta)$ (see Methods), and to four experimental traces, corresponding to four distinct cells, totaling 700 rotations \cite{Sowa:2005p8}.
According to our model, forward and backward steps across the same barrier should have the same average size.
In the simulations (Fig.~\ref{fig:positionmatters}B), mean forward and backward steps across the same barrier were found to be linearly correlated, though with a systematic offset toward smaller backward steps.
(As discussed above, this offset is likely the result of a bias in the step detection algorithm.)
We found that mean forward and backward step sizes across the same barrier are also positively correlated in the experiment (Fig.~\ref{fig:positionmatters}D), in agreement with our prediction (with the same bias towards smaller backward steps).
Overall, these results suggest that the absolute position of the rotor accounts for much of the variability observed in step size, and supports our model of a fixed, nearly periodic contact potential.

We next show that, both in the simulation and in the experiment, the barriers with a high frequency of backward steps are the same barriers where step sizes are short.
To this end we plot the frequency ratio of backward steps to forward steps across each barrier versus the average forward step size (which we showed in Fig.~\ref{fig:positionmatters}B,D correlates with the backward step size), for both simulations (Fig.~\ref{fig:positionmatters}C) and experiment (Fig.~\ref{fig:positionmatters}E). 
In both cases backward-step frequencies fall off sharply with average forward step size.

Since barriers where steps are smaller have higher backward-step frequencies than other barriers, they contribute more to the average backward step size. Therefore, the mean backward step is smaller than $360\degree/26\approx 13.8\degree$, which would be the mean backward step size if all barriers contributed equally. In contrast, forward-step frequencies vary little from barrier to barrier, so that all barriers contribute more or less equally to the average forward step size, which is therefore approximately $13.8\degree$. This explains why backward steps are smaller than forward steps on average.

\begin{figure*}
\includegraphics[width=\linewidth]{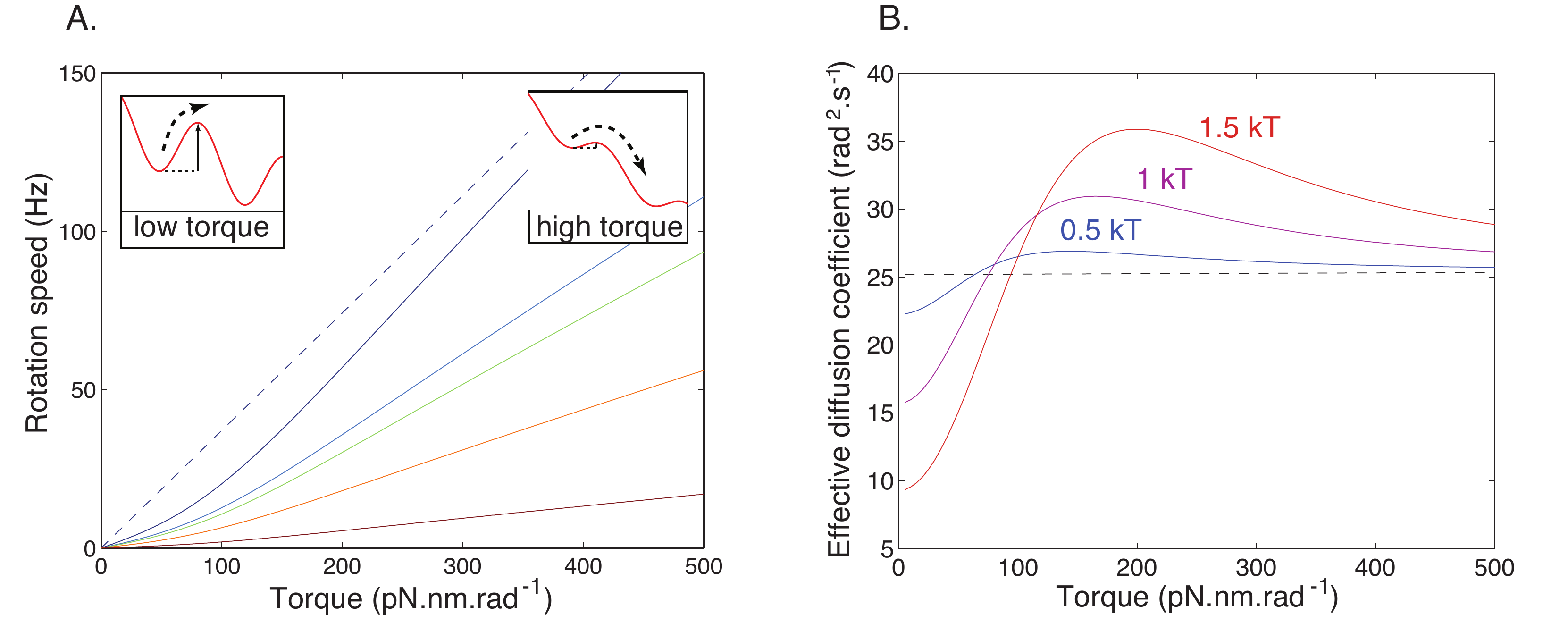}
\caption{{\bf The model predicts a sublinear torque-speed relation and a peak in rotor diffusion. A.} Rotation speed of the rotor as a function of torque for loads with different drag  taken from \cite{Ryu:2000p12}: solid curves, from top to bottom, $2\pi\nu=2.7$, $4.3$, $5.1$, $8.5$, and $28$ pN.nm.s.rad${}^{-1}$. The potential was chosen to be perfectly 26-fold periodic: $V(\theta)=A\sin(26\theta)$, with $A=1.5\,kT$.
As torque increases, the rotation speed asymptotes to the behavior expected in the absence of barriers, $f = \tau/2\pi\nu$, represented here for $2\pi\nu=2.7$ pN.nm.s.rad${}^{-1}$ (dashed line). At small torques, the rotation speed is limited by the rate of barrier crossing (left inset), while at high torques the tilt makes barriers easy to cross (right inset), and rotation is only limited by drag.
{\bf B.} Effective diffusion coefficient as a function of torque for a load with drag coefficient $2\pi\nu=1$  pN.nm.s.rad${}^{-1}$: solid curves, $V(\theta)=A\sin(26\theta)$ with $A=0.5\, kT$, $1\, kT$, and $1.5\, kT$. The real diffusion coefficient $D$ is represented by a dashed line.
}
\label{fig:tauvf}
\end{figure*}

\medskip\noindent\textbf{Sublinear torque-speed relation.}
The recognition that steps are barrier-crossing events has a direct implication for how rotation speed depends on torque. In the absence of contact forces ({\em i.e.} $V(\theta)=0$), the average rotation speed $f$ depends linearly on torque: $f=\tau/2\pi\nu$.
However, when the contact forces are comparable to the torque, rotation is hindered by barriers, and the system spends much of the time in local energy minima. Rotation is then not only limited by drag, but also by the rate of barrier crossing, leading to lower rotation speeds: $f<\tau/2\pi\nu$. We computed analytically (see Methods) the torque-speed relation for loads with various drag coefficients for a perfectly $26$-fold periodic sinusoidal potential with amplitude $1.5kT$, as shown in Fig.~\ref{fig:tauvf}A (using approximately $26$-fold periodic potentials yielded qualitatively identical results).
At high torques, the linear relation is recovered asymptotically,
which follows because increasing torque decreases the barriers to forward rotation, and eventually eliminates them completely, as shown in the insets to Fig.~\ref{fig:tauvf}A.

\medskip\noindent\textbf{Rotor diffusion.}
Fig.~\ref{fig:tauvf}B shows the effect of the contact potential on the effective long-time diffusion coefficient $D_{\rm eff}$. At low torques, diffusion is slowed down by barriers, while at high torques one recovers the natural diffusion coefficient $D$. Interestingly, at intermediate torques rotor diffusion is actually enhanced by the contact potential. In this regime, the contact potential is a small but variable correction to the torque. This variability contributes to the variance of the rotation speed, thus effectively enhancing rotor diffusion (see SI Text).  At large torques, $D_{\rm eff}$ approaches $D$ asymptotically:
\begin{equation}
D_{\rm eff}=D\left[1+3\frac{\langle (\partial_{\theta}V)^2\rangle_{\theta}}{\tau^2}\right].
\end{equation}

To check the consistency of the predicted diffusive behavior against previous results, we compared our model's prediction with experimental measurements of the variance in the rotation time \cite{Samuel:1996p40,Samuel:1995p16}. In \cite{Samuel:1996p40}, the rotation time of a tethered cell was measured. Simple diffusion predicts that the variance in rotation time per cycle is:
\begin{equation}
\<\delta T^2\>=\frac{2D}{(2\pi)^2 f^3}.
\end{equation}
For a single fully-powered torque-generating unit, torque is estimated to be $\tau=250$ pN.nm.rad${}^{-1}$ \cite{Ryu:2000p12}. The measured speed in \cite{Samuel:1996p40} was noisy and depended on the particular cell but was about $f\approx 3-5$ Hz for three torque-generating units, leading to an estimate for the drag coefficient $\nu=\tau/(2\pi f) \approx 25-40$ pN.nm.s.rad${}^{-2}$, and for the diffusion constant $D=kT/\nu \approx 0.1-0.15$ rad${}^{2}$/s. Thus the variance in cycle time for three torque-generating units predicted by simple diffusion is $\<\delta T^2\>\approx 0.7-2.0\cdot 10^{-4}$ s${}^2$, which is consistent with the reported value of $\<\delta T^2\>\approx 2.1\cdot 10^{-4}$ s${}^2$ \cite{Samuel:1996p40}. We conclude that it may not be necessary to consider other sources of fluctuations ({\em e.g.} proton translocations) to explain the observed variance in cycle time. Note that in these experiments the torque was high, and therefore contact forces are not expected to have had a significant effect on diffusion.

Our model also predicts a negative feedback reaction from the MotA/B protein springs that in principle could reduce diffusion. Namely, every time the rotor moves forward the springs relax causing a transient decrease of torque, and the opposite every time the motor moves backward. To estimate the magnitude of this effect, we model torque dynamics by linking spring elongation to rotor position, and by assuming that the springs ``restretch'' to their equilibrium position prescribed by the PMF with a characteristic relaxation time (see Methods). Within this model, the effective diffusion coefficient is found to be
$D(1-2\mu/\nu)$,
where $\mu$ is the slope of the torque-speed relation of the motor near stall. The value of $\mu$ ranges from $.03$ to $.07$ pN.nm.s.rad${}^{-2}$ depending on the temperature \cite{Chen:2000p5} and is therefore much smaller than relevant values of the drag coefficient $\nu\approx 0.5 - 50$ pN.nm.s.rad${}^{-2}$. We conclude that in the conditions of the discussed experiments, the effect of negative feedback from the springs on diffusion is negligible.

\medskip\noindent\textbf{Distribution of waiting times.}
According to our model, the distribution of waiting times between steps is expected to be roughly exponential. The distribution of waiting times between steps (forward or backward) for an exactly 26-fold periodic potential is indeed exponential (Fig.~S2A). When the potential is heterogeneous, the average waiting time depends on the barrier. Even though the distribution of waiting times across each barrier is exponential, the overall waiting-time distribution is not, appearing rather as a ``stretched'' exponential (Fig.~S2B). The experimental distribution also resembles a stretched exponential (Fig.~S2C).

\section*{Discussion}
Our model explains stepping of the bacterial flagellar motor by 
interpreting its rotation as a viscously damped random walk driven by a constant torque and by a heterogeneous contact potential caused by the physical irregularities of the rotor. In this picture, steps are recognized as barrier-crossing events between adjacent minima of a tilted and corrugated energy potential. Corrugations are caused by contact between the stators and the protein arrays (FliG, among others) making up the rotor structure. Recently a more accurate picture of this structure has emerged, thanks notably to electron microscopy studies \cite{Thomas:2006p44}.

Our model predicts a $2\pi$ periodicity of the potential, so that the absolute angular position of the rotor with respect to the stator is an underlying determinant for step statistics, and this prediction is found to be consistent with the available experimental data. In particular, our model offers an explanation for the experimental observation that backward steps are smaller than forward steps on average.

Another prediction of the model is that rotor speed grows sublinearly as a function of torque. At low torques rotation is slow because of trapping in local minima, whereas at high torques the barriers between minima are lowered and eventually eliminated. Additionally, we predict that at low torques rotor diffusion is hindered by barriers, while at high torques the variability of the potential actually enhances diffusion. Although in principle other sources of fluctuations, such as ion translocation, could impact rotor diffusion, we showed that in the relevant regimes simple diffusion can account for nearly all of the observed variance in cycle time \cite{Samuel:1996p40}.
In order to verify these predictions experimentally, one would need to simultaneously measure the rotor speed and the proton (or Na${}^{+}$) motive force, believed to be proportional to torque at low speeds, in the regime where torque and contact forces are comparable. (Note that in the stepping data \cite{Sowa:2005p8} we have analyzed, torque could vary during the course of the experiment as the result of changes in the number of motor units.)

Our model is consistent with other experimental results on the bacterial flagellar motor. Because the model relies on the assumption that the energy from ion translocation is reversibly stored in protein springs \cite{MEISTER:1989p20}, it implies a near-perfect efficiency of the motor at low torques \cite{MEISTER:1987p19}. The same mechanism can account for both clockwise and counterclockwise motor rotation \cite{BlairBerg88}---these two cases simply corresponding to the springs being stretched in opposite directions. If the contact potential stays the same when the motor changes direction, our model predicts that backward steps will occur preferentially at the same absolute angles irrespectively of the direction of rotation. 
The observation that the duty ratio is very close to one even with a single torque-generating unit \cite{Ryu:2000p12} can be encompassed in our model by assuming that each torque-generating unit comprises at least two springs.

In our analysis we have neglected one effect that is not crucial for our analysis, but which may prove important for inferring the detailed nature of the contact potential. Specifically, we have assumed that equilibration of the elastic linkage between the motor and the load is rapid compared to the waiting time between steps. For a torsion constant $k_{\theta}\approx 400$ pN.nm.rad${}^{-2}$ \cite{BLOCK:1989p29,Block:1991p37} between the rotor and the load, and a drag coefficient $2\pi\nu\approx 1.0-10$ pN.nm.s.rad${}^{-2}$, the relaxation time is $t_{\theta}=\nu/k_{\theta}=0.4-4$ ms. In contrast, the typical waiting time between steps ranges from $10$ to $100$ ms, depending on experimental conditions. If the elastic linkage was too soft, the polystyrene bead would respond to the motion of the rotor with a delay $t_{\theta}$, and steps would be smoothed out. This does not seem to occur in the experiment.

Another effect, which we have considered (see Rotor diffusion) but did not include in our simulations, is the relaxation of MotA/B protein springs as a rotor step occurs.  For example, when the rotor moves forward, the torque decreases because the protein springs relax. Usually these springs are restretched so quickly by ion translocation that the transient decrease of torque can be neglected. However, during a barrier crossing event the rotor motion might be so fast that protons are not able keep up. This would result in a temporary drop in torque and make barrier crossing more difficult. A similar argument applies to backward steps. We have already shown that at the ``mean-field'' level, where rotation speed and ion flux are time-averaged, this negative feedback has only a small effect. However, the instantaneous rotation speed during a step can be much larger than its average. 
How fast can a proton translocate through the motor? The maximum flux of protons through a single motor unit can be estimated by considering the maximum rotation speed before the torque starts dropping (the ``knee'' of the torque-speed relationship \cite{Berg:1995p18}). For a single motor unit in natural conditions, this speed is about 150 Hz for a torque of 250 pN.nm.rad${}^{-1}$ \cite{Ryu:2000p12}. The power generated by the motor is then $150\times 2\pi\times$ 250 $ \approx 240,000$ pN.nm.s${}^{-1}$. Each proton provides at most $150$ mV $\times e = 24$ pN.nm, so that the number of protons per second is at least $240,000/24=10,000$. The timescale of proton passage is therefore less than $0.1$ ms. A single rotor step corresponds to the passage of $\approx 3$ protons \cite{Sowa:2005p8}, so restretching the protein springs should take less than $0.3$ ms, which is below the current experimental time resolution. For comparison, an instanton calculation \cite{WentzelFriedlin98} reveals that the typical time for crossing a barrier is bounded from below by $(2\pi/26)\times \nu / \max |\partial_\theta U| \approx 0.2$ ms (for $2\pi\nu=1$ pN.nm.s.rad${}^{-1}$, and $U(\theta)/kT=1.5\cos(26\theta)-10\theta$).

Other molecular motors have shown stepping behavior, including the actin-myosin motor \cite{Finer:1994p34}, the dynein-microtubule motor \cite{Mallik:2004p30}, and kinesin \cite{AsburyFehr03}. In these ATP-powered motors, which are less powerful than the bacterial flagellar motor by orders of magnitude, stepping is a built-in and essential part of motor operation. By contrast, we have argued that in the bacterial flagellar motor the observed stepping arises solely from steric hindrance.

Our work leaves open a number of questions. It would be interesting to infer the precise form of the contact potential $V(\theta)$ from rotation data and see how and whether it varies in time and among motors. To this end a more sophisticated approach to learning the potential may be required, {\em e.g.} employing maximum likelihood techniques. Lastly, one still needs to understand the mechanism of torque generation, including the role played by the discreteness of ion translocation, the chemical nature of protein springs and their attachment sites, as well as the energy conversion process.

\section*{Methods}

All the experimental data presented in this paper is from \cite{Sowa:2005p8} and were used with the kind permission of Richard Berry.

The simulation data were obtained by numerical integration of Eq.~\ref{eq:langevin} by Euler's method.
The same step finding algorithm as the one described in \cite{Sowa:2005p8} was used on both simulation data and experimental data to extract steps.

The step frequencies presented in Fig.~\ref{fig:positionmatters}A were obtained analytically using first-passage theory \cite{GardinerBook,LindnerKostur01}:
Label the wells $i=1,\ldots,26$, and denote the local minima of $U(\theta)$ by $\theta_i$.
Consider three consecutive wells centered at $\theta_{i-1}$, $\theta_i$ and $\theta_{i+1}$ respectively. Starting at $\theta_i$, call $\pi_i^+$ the probability of first jumping forward and $\pi_i^-$ the probability of first jumping backward. These probabilities are given by:
\begin{equation}\label{eq:pi}
\frac{\pi_i^+}{\pi_i^-}=\frac{\int_{\theta_{i-1}}^{\theta_i} d\theta\,e^{U(\theta)/kT}}{\int_{\theta_i}^{\theta_{i+1}} d\theta\,e^{U(\theta)/kT}},\quad \pi^+_i+\pi^-_i=1.
\end{equation}
Given the transition probabilities $\pi_i^+$ and $\pi_i^-$, we write a master equation for the probability $p_i(m)$ of the rotor being in well $i$ after $k$ steps.
\begin{equation}
p_i(m)=p_{i-1}(m-1)\pi_{i-1}^++p_{i+1}(m-1)\pi_{i+1}^-.
\end{equation}
At large $m$ a stationary state is reached, and $p_i:=p_i(m\to \infty)$ satisfies the conservation equation:
\begin{equation}
p_i=p_{i-1}\pi_{i-1}^++p_{i+1}\pi_{i+1}^-.
\end{equation}
The step frequencies presented in Fig.~\ref{fig:positionmatters} are then given by $p_i\pi_i^+$ and $p_i\pi_i^-$. (Note that the step frequencies sum to one, $\sum_{i=1}^{26} p_i\pi_i^++\sum_{i=1}^{26}p_i\pi_i^-=1$.)

The torque-speed relation shown in Fig.~\ref{fig:tauvf}A was also estimated using first-passage theory. For simplicity we assumed a perfectly 26-fold periodic potential, but the results are qualitatively the same when the periodicity is only approximate. 
The average time to move from one minimum of the potential to the next is given by:
\begin{equation}
\<t_{\rm step}\>=\frac{1}{D}\frac{\int_{0}^{\Delta}d\theta\,e^{-U(\theta)/kT}\int_{\theta}^{\theta+\Delta}d\theta'\, e^{U(\theta')/kT}}{1+e^{-\tau\Delta/kT}},
\end{equation}
where $\Delta=2\pi/26$ and $\tau$ is the torque. This yields a rotation speed:
\begin{equation}\label{eq:omega}
f=\frac{\Delta(\pi^+-\pi^-)}{2\pi \<t_{\rm step}\>}=\frac{\Delta}{2\pi \<t_{\rm step}\>} \tanh\left(\frac{\tau\Delta}{2kT}\right),
\end{equation}
where $\pi^+$ and $\pi^-$ are obtained from Eq.~\ref{eq:pi}.
The asymptotic expansion of Eq.~\ref{eq:omega} for $\tau\gg \left|\partial V/\partial \theta\right|,\,kT$ yields
\begin{equation}\label{eq:asymp}
f \approx \frac{\tau}{2\pi\nu}\left[1-\frac{\langle (\partial_{\theta}V)^2\rangle_{\theta}}{\tau^2}\right].
\end{equation}
The effective diffusion coefficient $D_{\rm eff}$ shown in Fig.~\ref{fig:tauvf}B is estimated using similar techniques ({\em cf.} \cite{GardinerBook,LindnerKostur01} and SI Text).

To estimate the magnitude of the negative feedback of the MotA/B protein springs on diffusion, we model the dynamics of rotor angle and torque by the following mean-field differential equations:
\begin{eqnarray}
\frac{d\theta}{dt} & = & \frac{\tau}{\nu} + \xi(t), \label{eq:torque}\\
\frac{d\tau}{dt} &=& - \frac{\tau - \tau_{\textsc{pmf}}}{t_{\rm relax}} - k \frac{d\theta}{dt}.\label{eq:angle}
\end{eqnarray}
The first equation is the same as Eq.~\ref{eq:langevin} but with a variable torque and without contact forces. In the second equation, torque is assumed to follow the stretching/unstretching of the springs with rotation, and therefore the rate  of change of torque is linearly related to the rate of change of rotor angle through an effective spring constant $k$ (second term of r.h.s. of Eq.~\ref{eq:angle}). At the same time, due to the restretching of springs upon proton passage, torque relaxes toward its equilibrium value $\tau_{\textsc{pmf}}$ (first term of r.h.s. of Eq.~\ref{eq:angle}). Solving the second equation for steady-state rotation yields the torque-speed relationship for the motor:
$\tau=\tau_{\textsc{pmf}}- 2\pi \mu f$, with $\mu:=kt_{\rm relax}$.
Solving both equations for the effective diffusion coefficient in the limit $\mu/\nu\ll 1$, we find $\<\delta\theta^2\>/t = 2D\left[1-2\mu/\nu+O((\mu/\nu)^2)\right]$.

\medskip

For all numerical simulations and analytic calculations the potential was chosen to be of the form:
\begin{equation}
V(\theta)=A\cos(26\,\theta) + B\cos(10\,\theta)+C\cos(11\,\theta).
\end{equation}
Experimentally, there is evidence for components of the contact potential with $\sim 10-11$ fold periodicity (see Ref.~\cite{Sowa:2005p8}, Fig. 3b).
Except where stated otherwise, we used $A=1.5 \,kT$, $T=290\,{\rm K}$ and $2\pi\nu=1$ pN.nm.s.rad${}^{-1}$. 
In Fig.~\ref{fig:backstepsaresmaller} and \ref{fig:positionmatters}A, we used $B=C=0.6\,kT$. In Fig.~\ref{fig:positionmatters}B and \ref{fig:positionmatters}C we chose three sets of values for $B$, $C$, and $\tau$, in units of $kT$: ($B=0.6,\ C=0.6,\ \tau=10$), ($B=1,\ C=0.5,\ \tau=15$) and ($B=0.3,\ C=0.8,\ \tau=12$). In Fig.~\ref{fig:tauvf}, the torque-speed relation and rotor diffusion were calculated with $B=C=0$.

\section*{ACKNOWLEDGMENTS} We thank Richard Berry, Avigdor Eldat, Yigal Meir, Teuta Pilizota, Anirvan Sengupta, and Aleksandra Walczak for helpful suggestions. T.M. was supported by a Human Frontier Science
Program fellowship, and N.S.W.  by National Institutes
of Health Grant R01 GM082938.

\bibliographystyle{unsrt}
\bibliography{bfm,bfm2}

\end{document}


\title{Supporting Information}


\maketitle

\medskip\noindent\textbf{Step finding algorithm.}
Steps were found using the algorithm described in \cite{Sowa:2005p8}. First, an entire episode (angular position {\em vs.} time) was fitted by a single step function, and thus divided into two intervals. This procedure was repeated iteratively $N_{\rm steps}$ times---at each iteration the interval for which the data had the largest range of angles was replaced by a best-fit step function. Then, a quality factor $Q=({\bar x_2}-\bar{x_1})^2/({\rm var}\  x_1/n_1+{\rm var}\  x_2/n_2)$ was calculated for each assigned step, where $x_{1}$ ($x_2$) is the angular position left (right) of the step, and $n_{1}$ ($n_2$) is the number of data points on the left (right) of the step. Finally, steps with a low quality factor were removed and their adjacent intervals merged until all steps have a quality factor greater than $Q_{\rm min}$.

This step finding algorithm introduces a small bias that tends to underestimate the sizes of backward steps. For each interval, the algorithm finds the mean value of the angular position in this interval.
But steps do not occur instantaneously, and the data points leading from one interval to the other are themselves included in the interval means. As a result, these means are biased toward where the rotor is coming from and where it is going. When the rotor is stepping forward, these two biases tend to cancel each other. However, when the rotor steps backward, these steps are usually both preceded and followed by forward steps. Thus, the interval before the backward step is biased toward lower angular positions, and the interval after the backward step is biased toward higher positions. As a result, the biases reinforce each other such that backward steps are estimated to be smaller than forward steps. We have checked that even on a perfectly 26-fold periodic potential $V(\theta)=A\cos(26\theta)$, with $A=1.5 kT$ and a torque $\tau=10 kT$, the most probable forward step size according to the algorithm was $13.8\degree\approx 360\degree/26$, which is correct, while the most probable backward step size was $13.0\degree$, which is $0.8\degree$ to small.

\medskip\noindent\textbf{Effective diffusion.}
The effective diffusion coefficient is estimated using the techniques of first-passage theory \cite{GardinerBook}. For an exactly 26-fold periodic potential, this effective diffusion coefficient is given by \cite{LindnerKostur01}:
\begin{equation}\label{eq:Deff}
D_{\rm eff}=\frac{\Delta^2}{2}\left[\frac{1}{\<t_{\rm step}\>}+\frac{(\pi^+-\pi^-)^2}{\<t_{\rm step}\>^3}(\<t_{\rm step}^2\>-2\<t_{\rm step}\>^2)\right],
\end{equation}
where $\Delta=2\pi/26$ is the step size. $\pi^+$ and $\pi^-$ are the probabilities of first jumping forward or backward, respectively, and are given by:
\begin{equation}\label{eq:pi}
\pi^+-\pi^-=\tanh\left(\frac{\tau\Delta}{2kT}\right),\quad \pi^++\pi^-=1.
\end{equation}
$t_{\rm step}$ is the time before a step occurs (forward or backward). Its first two moments are:
\begin{equation}
\<t_{\rm step}\>=\frac{1}{D}\frac{\int_{0}^{\Delta}d\theta\,e^{-U(\theta)/kT}\int_{\theta}^{\theta+\Delta}d\theta'\, e^{U(\theta')/kT}}{1+e^{-\tau\Delta/kT}},
\end{equation}
\begin{equation}
\<t_{\rm step}^2\>=\frac{2}{D(1+e^{-\tau\Delta/kT})}\int_{0}^{\Delta}d\theta\,e^{\frac{U(\theta)}{kT}}\int_{\theta-\Delta}^{\theta}d\theta'\, e^{-\frac{U(\theta')}{kT}}t_1(\theta')
\end{equation}
with
\begin{equation}
\begin{split}
t_1(\theta)&=\frac{\left(\int_{-\Delta}^{\theta}d\theta'\, e^{U(\theta')/kT}\right)\int_{\theta}^{\Delta}d\theta'\,\int_{-\Delta}^{\theta'}d\theta''\,e^{[U(\theta')-U(\theta'')]/kT}}{D\int_{-\Delta}^{\Delta}d\theta'\,e^{U(\theta')/kT}}\\
&-\frac{\left(\int_{\theta}^{\Delta}d\theta'\, e^{U(\theta')/kT}\right)\int_{-\Delta}^{\theta}d\theta'\,\int_{-\Delta}^{\theta'}d\theta''\,e^{[U(\theta')-U(\theta'')]/kT}}{D\int_{-\Delta}^{\Delta}d\theta'\,e^{U(\theta')/kT}}.
\end{split}
\end{equation}
Expanding Eq.~\ref{eq:Deff} for large torques gives:
\begin{equation}
D_{\rm eff}=D\left[1+3\frac{\langle (\partial_{\theta}V)^2\rangle}{\tau^2}\right].
\end{equation}

\medskip\noindent\textbf{Waiting-time distribution.}
We recorded the distribution of waiting times obtained by the step-finding algorithm, both for the simulation and for the experiment. For the simulation, in the case of a perfectly 26-fold periodic potential (Fig.~S2A), there is only one type of barrier, and the waiting-time distribution is approximately exponential. When the potential is only approximately 26-fold periodic (Fig.~S2B), the waiting-time distribution is the sum of 26 exponentials, and resembles a stretched exponential. The waiting-time distribution for the experiment (one cell, Fig.~S2C) is consistent with a sum of exponentials. The stretched appearance of the experimental distribution may be due to non-uniform barriers, as predicted by our model, but may also be due in part to the observed variability in average speed, presumably due to changes in torque, during the course of the experiment.

\begin{figure*}[p]
\includegraphics[width=.5\linewidth]{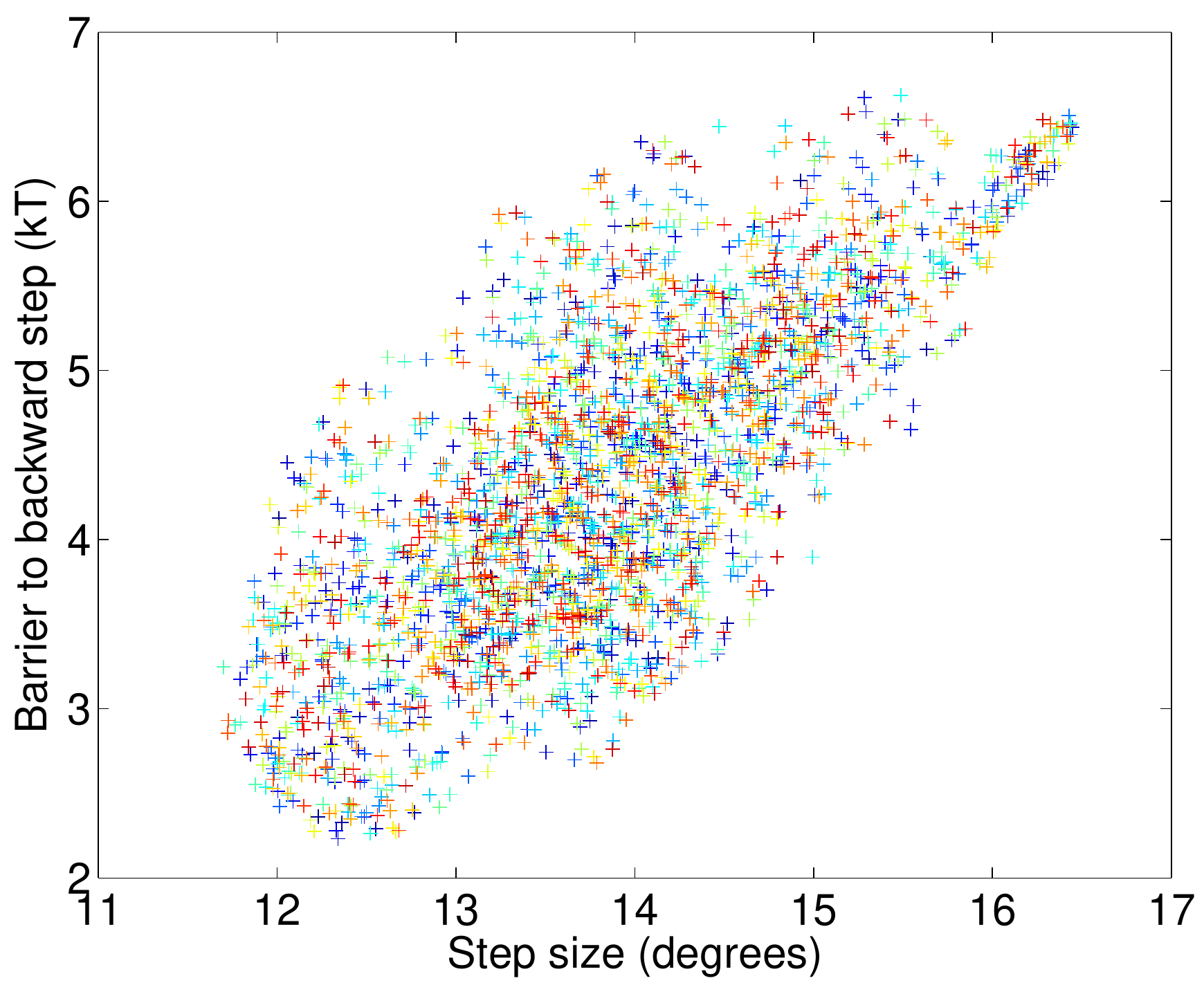}
\caption{\sf {\bf The height of barriers to backward steps is positively correlated with the backward step size.} For 100 randomly generated potentials $V(\theta)$, we plotted the size of each of the 26 steps versus the height of the corresponding barrier to backward steps ({\em i.e.} the energy difference between the minimum of $U(\theta)$ and its immediately preceding maximum). Each color corresponds to a particular potential.
We used: $V(\theta)=A_{26}\cos(26\theta)+A_{6}\cos(6\theta)+A_{9}\cos(9\theta)+A_{10}\cos(10\theta)+A_{11}\cos(11\theta)+A_{16}\cos(16\theta)$, with $A_{26}=1.5 kT$, and $A_6,\ldots,A_{16}$ drawn uniformly at random with fixed total power $\sum_j A_j^2 = (1.0 kT)^2$. The torque is $\tau=10kT\approx 41$ pN.nm.rad${}^{-1}$.
}
\label{fig:DeltaUvsDeltatheta}
\end{figure*}

\begin{figure*}[p]
\includegraphics[width=.32\linewidth]{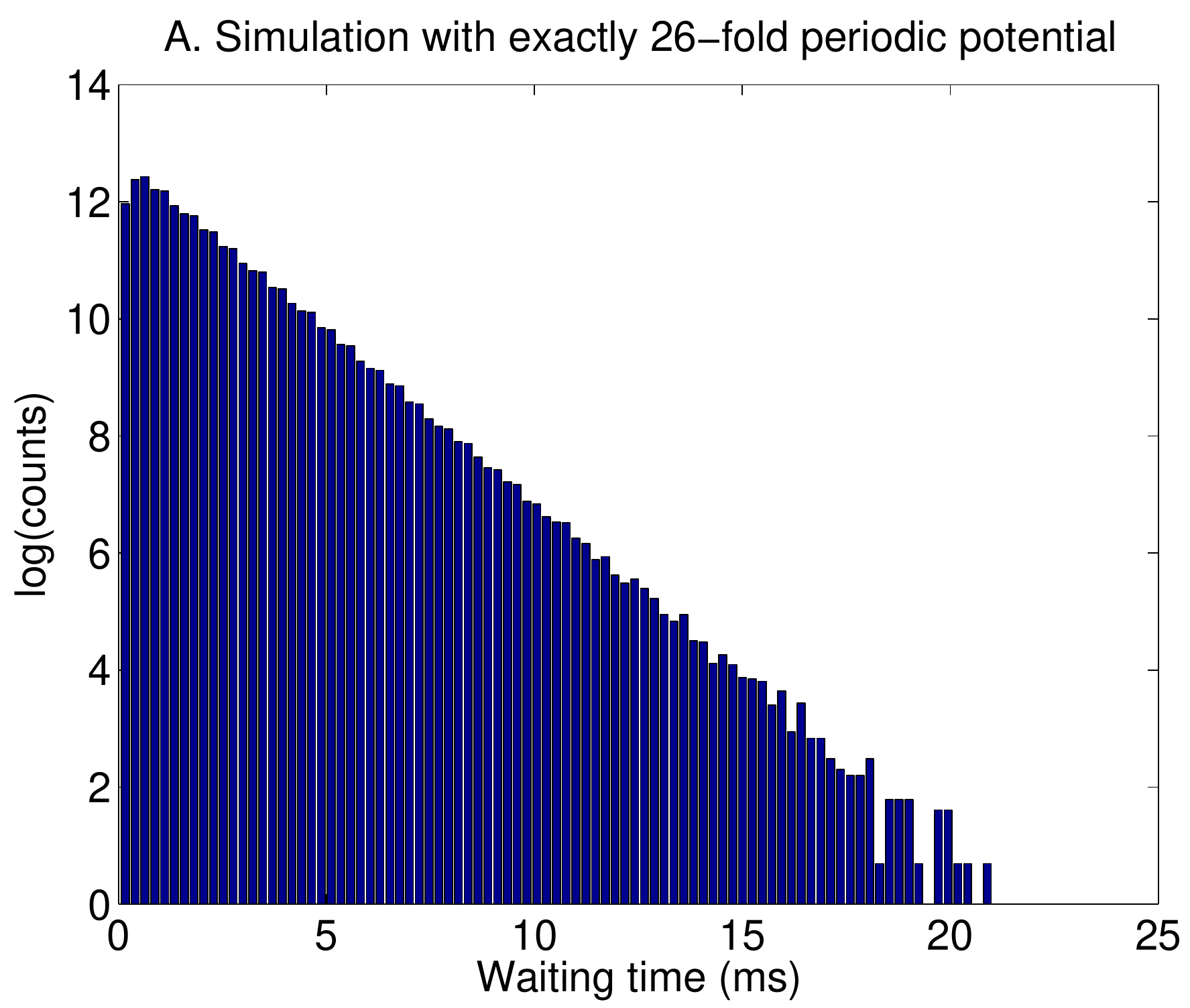}
\includegraphics[width=.32\linewidth]{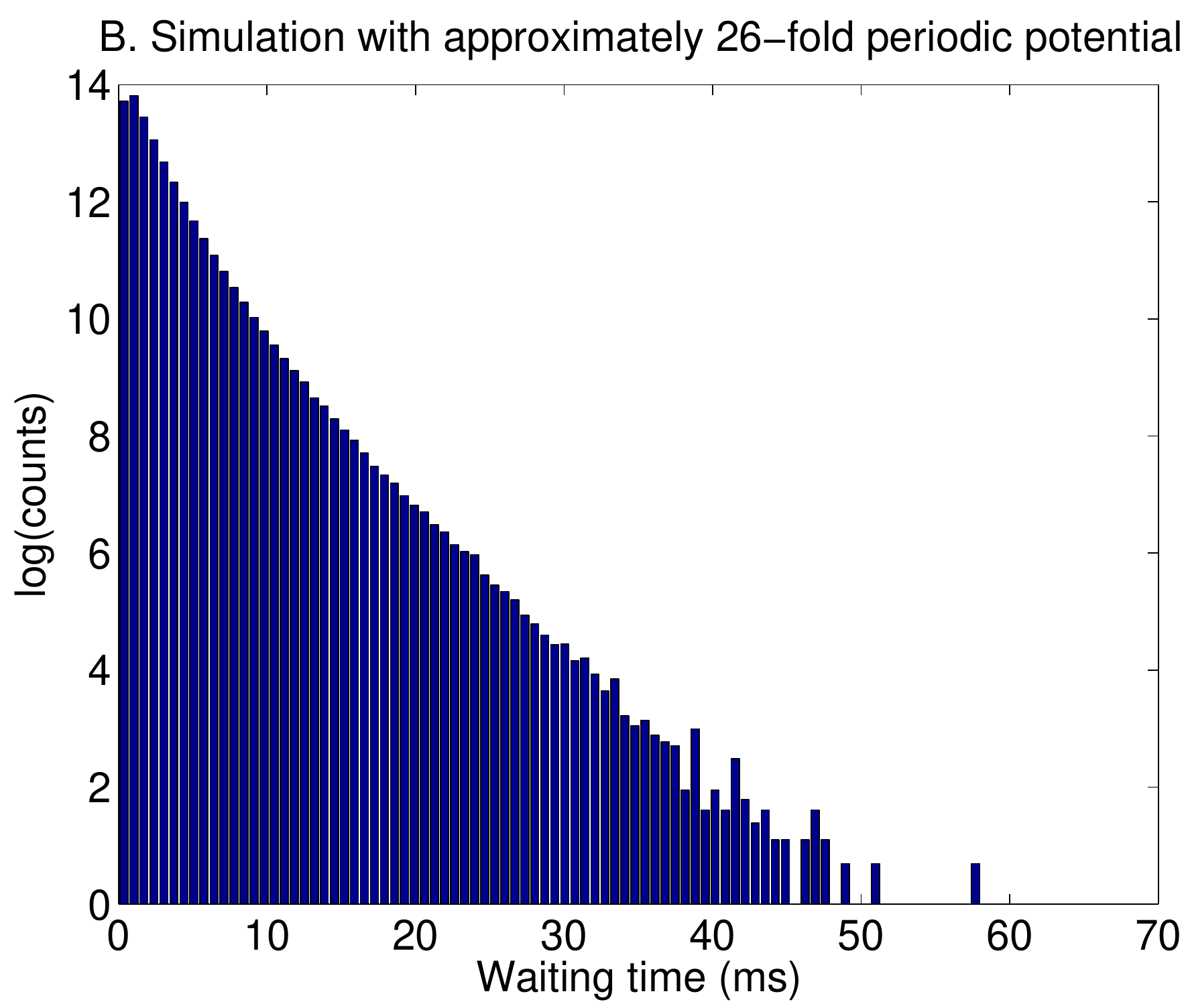}
\includegraphics[width=.32\linewidth]{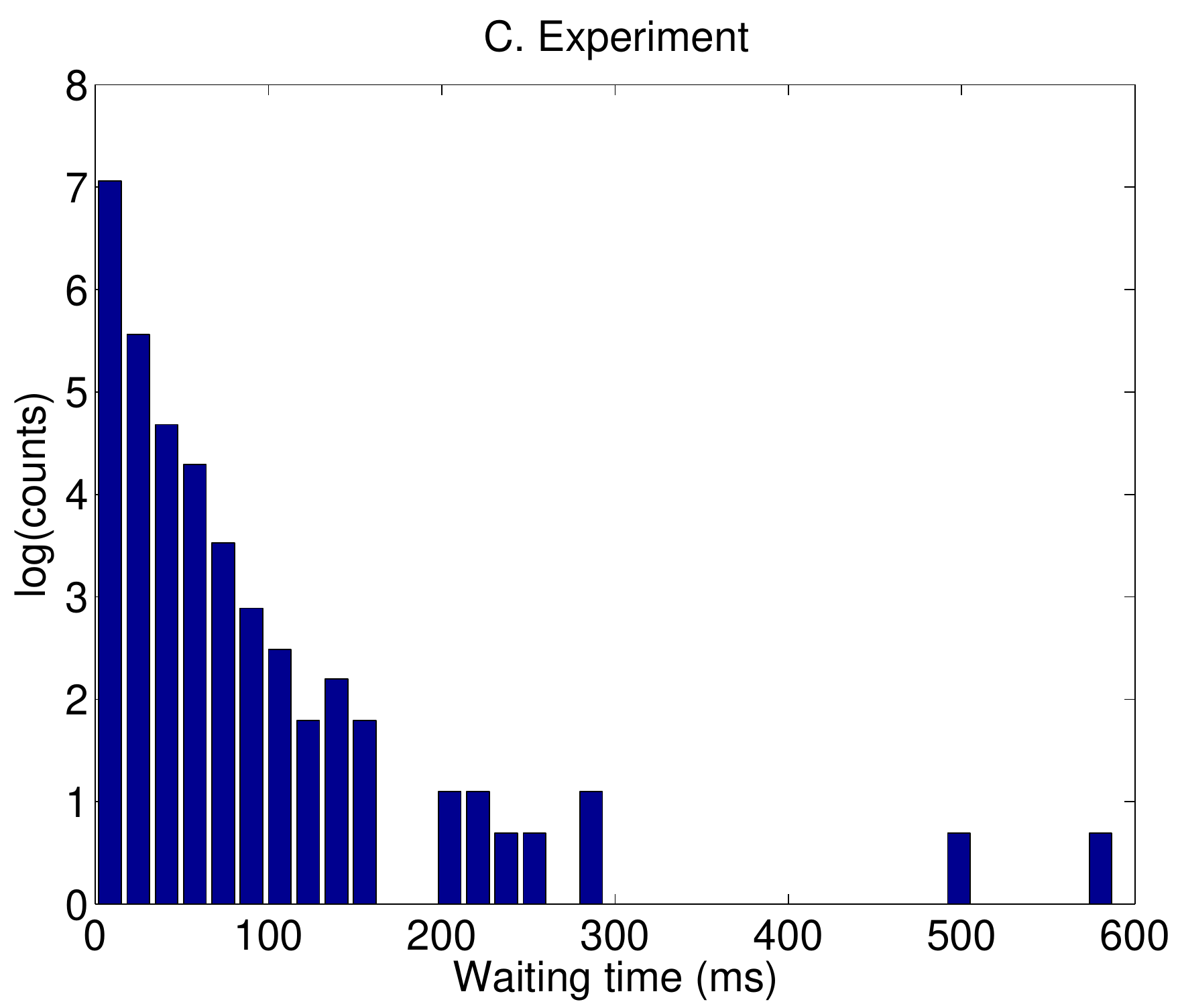}
\caption{\sf {\bf The distribution of waiting times between steps (backward or forward). A.} Simulation with a perfectly 26-fold periodic contact potential $V(\theta)$. There is only one type of barrier, and the distribution of waiting times is roughly exponential. (The contact potential is $V(\theta)=A\sin(26\theta)$ with $A=1.5kT$, and torque $\tau=10kT\approx 41$ pN.nm.rad${}^{-1}$.) {\bf B.} Simulation with an approximately 26-fold periodic contact potential. There are 26 types of barriers, each of them having a different characteristic time. The overall waiting time distribution is a therefore the sum of 26 exponentials. (The contact potential and torque are the same as in Fig.~2 of the main text.) {\bf C.} Experiment\,-- the distribution is consistent with a sum of exponentials.
}
\label{fig:DeltaUvsDeltatheta}
\end{figure*}

\bibliographystyle{unsrt}
\bibliography{bfm,bfm2}